\documentclass[sigconf, nonacm]{acmart}

\newcommand\vldbdoi{}

\newcommand\vldbavailabilityurl{}

\usepackage{balance}  
\usepackage{amsmath}

\usepackage{gensymb}
\usepackage{url}
\usepackage{subfigure}
\usepackage{diagbox}

\usepackage{graphicx,epstopdf,url,color}
\usepackage{balance}
\usepackage{placeins}
\usepackage{amsthm}
\usepackage{threeparttable}

\usepackage{multirow}
\usepackage{bm}
\usepackage{algpseudocode}
\usepackage{amssymb}
\usepackage[lined,linesnumbered,boxed,commentsnumbered, ruled,vlined]{algorithm2e}
\usepackage{float}
\usepackage{array}
\newfloat{algorithm2e}{t}{lop}
\newcolumntype{C}[1]{>{\centering\let\newline\\\arraybackslash\hspace{0pt}}m{#1}}

\newcommand{\etc}{\emph{etc.}\xspace}
\newcommand{\eat}[1]{}

\usepackage{hyperref}
\let\svthefootnote\thefootnote
\newcommand\freefootnote[1]{%
  \let\thefootnote\relax%
  \footnotetext{#1}%
  \let\thefootnote\svthefootnote%
}

\usepackage{xcolor}

\begin{document}
\title{Three Pillars Towards Next-Generation Routing System [Vision]}

\author{Lei Li$^{1,2}$, Mengxuan Zhang$^3$, Zizhuo Xu$^2$, Yehong Xu$^2$, Xiaofang Zhou$^{2,1}$}
\affiliation{
\institution{
$^1$DSA Thrust \& INTR Thrust, The Hong Kong University of Science and Technology (Guangzhou), Guangzhou, China\\
$^2$Department of CSE, The Hong Kong University of Science and Technology, Hong Kong SAR, China\\
$^3$School of Computing, The Australian National Unversity, Australia
    }
    \country{}
}
\email{thorli@ust.hk, mengxuan.zhang@anu.edu.an, zxucj@connect.ust.hk, yxudi@connect.ust.hk, zxf@cse.ust.hk}
\begin{abstract}
The routing results are playing an increasingly important role in transportation efficiency, but 
they could generate traffic congestion unintentionally.
This is because the traffic condition and routing system are disconnected components in the current routing paradigm. 
In this paper, we propose a next-generation routing paradigm that could reduce traffic congestion by considering the influence of the routing results in real-time. Specifically, we 
regard the routing results as the root cause of the future traffic flow, which at the same time is identified as the root cause of traffic conditions. To implement such a system, we identify three essential components: 1) the traffic condition simulation that establishes the relation between traffic flow and traffic condition with guaranteed accuracy; 2) the future route management that supports efficient simulation with dynamic route update; 3) the global routing optimization that improves the overall transportation system efficiency. Preliminary design and experimental results will be presented, and the corresponding challenges and research directions will also be discussed.
\end{abstract}

\maketitle


\section{Introduction}
\label{sec:Introduction}
The intelligent transportation system is an essential infrastructure of modern smart cities because almost every production or living activity involves traveling from one place to another. As the core component of this grand system, the vehicle routing system determines how the traffic evolves. This is because, with the fast development of GPS and digital maps, more and more transportation relies on routing systems nowadays. Although the drivers are not required to strictly follow the navigation, the routing result still has an increasing impact on the driving behaviors. Moreover, as self-driving cars are fast developing, the impact of human behavior is decreasing while the fulfillment of navigation is increasing. Consequently, the routing system is playing a more and more important role in the transportation system.

In the past decades, routing system is also evolving fast from the perspectives of efficiency \cite{geisberger2008contraction,akiba2013fast,delling2017customizable,li2017experimental,ouyang2018hierarchy,li2019scaling,lakhotia2019planting,chen2021p2h}, accuracy \cite{kanoulas2006finding,ding2008finding,batz2009time,foschini2014complexity,li2017minimal,li2018go,li2019time,wang2019querying,li2020fastest,dan2023double,li2023efficient}, dynamicity \cite{malviya2011continuous,akiba2014dynamic,ouyang2020efficient,wei2020architecture,zhang2021efficient,zhang2021dynamic,zhangexperimental,zhang2023parallel,zhang2022relative}, scalability \cite{zhang2019efficient,li2020scaling,li2020fast,zhang2020stream,farhan2022batchhl,wang2021query,zhang2023universal}, and usability \cite{cao2012keyword,yang2014finding,wang2016effective,yawalkar2019route,liu2017finding,liu2022fhl,chondrogiannis2020finding,juttner2001lagrange,liu2021efficient,liu2023multi,liu2018finding,gong2019skyline,lu2020accelerating,chen2022constrained,zhang2021efficientConstrained,luo2022diversified,li2023finding,wang2023qhl}. Although these algorithms work on different problems with different techniques, they share the same routing paradigm: they only optimize one single query based on the predicted traffic. Such a paradigm has worked fine in the past, 
but its \textit{selfishness}\cite{roughgarden2002bad} has been growing and started to cause traffic congestion as more and more vehicles rely on navigation. Intuitively, suppose there are thousands of vehicles departing from the same place almost simultaneously and going to the same destination, then the current system would assign the same path for them such that congestion is inevitable. Even if they have different origins and destinations, congestion could still appear if their routes share some common road segments and flood into these segments at the same time. As a result, economic loss \cite{weisbrod2003measuring}, environmental \cite{armah2010systems} and health \cite{zhang2013air} issues would follow closely.

\begin{figure}[t!]
	\centering
	\includegraphics[width=3.2in]{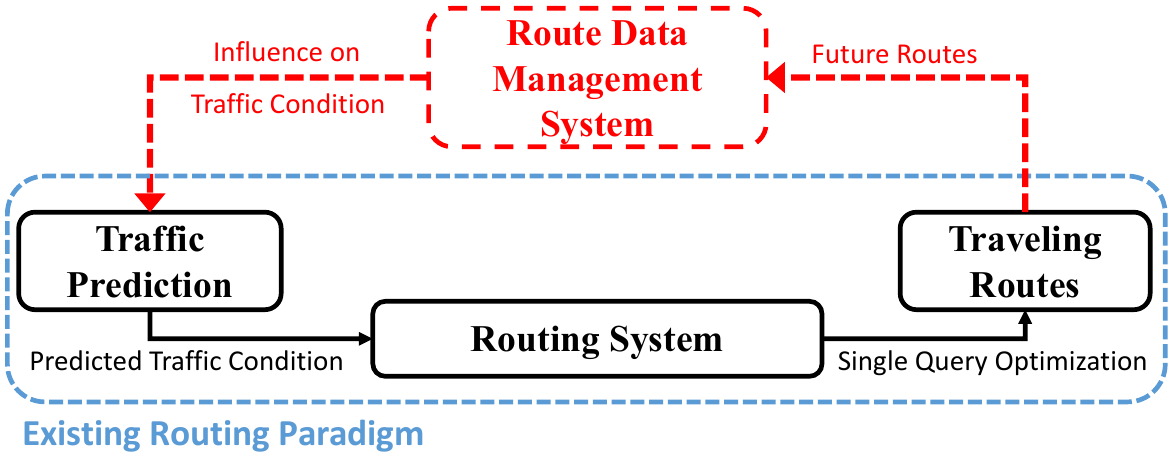}
	\caption{The Current and Future Routing Paradigms}
	\label{fig:Feedback}
\end{figure}

The reason for this phenomenon is the separation of \textit{traffic prediction system} and \textit{routing system}. Historically, the traffic prediction task is regarded as part of the spatiotemporal data analysis, while the routing is a part of graph and spatial data management. Specifically, on the one hand, the prediction task takes various historical, network, temporal, POI, and other side features as input, but ignores the effect of routing results. On the other hand, the routing algorithm only takes the predicted traffic as input and leaves the routing results with no further management. Such a disconnected one-way paradigm is illustrated in the blue box of Figure \ref{fig:Feedback}. 

In this paper, we propose a next-generation routing system to fix this long-existing problem by introducing a feedback line from the routing results back to the traffic prediction, as illustrated in red in Figure \ref{fig:Feedback}. However, it is non-trivial to achieve this, and we identify three critical components. \textbf{C1:} The first one is about correctness or accuracy since there is no direct relation between routing results and traffic prediction. The current prediction methods \cite{li2018multi,yuan2020effective,wang2019simple,zhang2018deeptravel,yang2013travel,hong2020heteta,hui2021trajnet,wu2020connecting,zheng2020gman,pan2019urban,yu2018spatio,lv2018lc,wu2021autocts,cirstea2022towards,jin2023spatio,wu2023autocts+,zhang2023mlpst} cannot take routing results as input, and they cannot scale to real-life large networks. Therefore, we first convert the routes into traffic flow, and then we propose to establish the relation between traffic flow and traffic conditions through small expert models. After that, we propose a macroscopic simulation to manage these models and routes to generate accurate traffic predictions based on routing results. It should be noted that such a paradigm would be the most robust model as it can adapt to any traffic pattern or network change timely and correctly.

\textbf{C2:} The second one is about simulation efficiency, which is an obstacle for real-time optimization. The existing simulation systems have to re-simulate all the traffic from the beginning whenever any route changes \cite{lopez2018microscopic,sumo2023official,waraich2015performance, matsim2023official,ramamohanarao2016smarts, xie2019generating, smarts2023official}. Obviously, it is time-consuming and makes the optimization impossible. To solve this problem, we first propose to regard these future routing schedules as trajectories and manage them accordingly. However, all the existing spatiotemporal systems only support retrieving and updating each trajectory data on its own but do not consider its influence on its spatiotemporal neighbors. Therefore, we propose a new future route management system \cite{zheng2015trajectory,zheng2011computing,wang2021survey,su2020survey,tao2001efficient,tao2001mv3r,tao2003tpr,ding2018ultraman,cudre2010trajstore,xie2017distributed,wang2015sharkdb} that supports fast trajectory insertion and deletion with its influence propagation such that no entire re-simulation is required for updates, and the simulation efficiency could be boosted.

\textbf{C3:} The third one is about the congestion reduction effectiveness or global traffic optimization. Most of the existing solutions are from the traffic assignments \cite{dafermos1980traffic,gugat2005optimal,kollias2021weighted,roughgarden2002bad,wardrop1952road,ban2008link}, which are slow to run, can only scale to toy network, and cannot utilize any new route outside the pre-defined ones. Another stream of \textit{global routing} research \cite{li2021towards,chen2020pay,arp2020dynamic} works on a similar problem. However, because they lack the simulation system, only the next edge of the current routing search could be optimized. So their effectiveness and efficiency are not satisfactory. To this end, we propose the global routing optimization framework that utilizes the previous two components to reduce traffic congestion by identifying the congested areas and re-routing their corresponding routes.

Our new proposed routing paradigm has a wider scope of applications beyond reducing congestion: 1) It can provide higher quality traffic prediction with detailed explanations to support traffic management decision-making; 2) It can evaluate network structure modification in real-time to support network optimization; 3) It is the core technique of smart city's digital twin to support real-time decision evaluation; 4) It is also the basis for other future transportation systems like low-altitude airspace for the unmanned aerial vehicles and conflict-free environment for the automated guided vehicles. The major contributions are summarized as: 

\begin{itemize}
\item[--] We propose a novel routing paradigm that integrates routing and traffic prediction together by identifying the root cause of traffic congestion in the current routing paradigm.
\item[--] We identify the key components to implement this paradigm and propose three pillar systems from the perspective of accuracy (traffic simulation), efficiency (future route management), and effectiveness (global routing optimization). 
\item[--] We provide the preliminary design and experimental results and discuss the future directions.
\end{itemize}


\section{Current Solutions and Limitations}
\label{sec:Related}
In this section, we briefly summarize the related topics and discuss why the current solutions cannot solve our problem.

\textbf{1) Route Planning}. 
The single-path methods \cite{geisberger2008contraction,akiba2013fast,delling2017customizable,li2017experimental,ouyang2018hierarchy,li2019scaling,lakhotia2019planting,chen2021p2h,kanoulas2006finding,ding2008finding,batz2009time,foschini2014complexity,li2017minimal,li2018go,li2019time,wang2019querying,li2020fastest,dan2023double,li2023efficient,malviya2011continuous,akiba2014dynamic,ouyang2020efficient,wei2020architecture,zhang2021efficient,zhang2021dynamic,zhangexperimental,zhang2023parallel,zhang2022relative,zhang2019efficient,li2020scaling,li2020fast,zhang2020stream,farhan2022batchhl,wang2021query,zhang2023universal,cao2012keyword,yang2014finding,wang2016effective,yawalkar2019route,liu2017finding,liu2022fhl,chondrogiannis2020finding,juttner2001lagrange,liu2021efficient,liu2023multi,liu2018finding,gong2019skyline,lu2020accelerating,chen2022constrained,zhang2021efficientConstrained,luo2022diversified,li2023finding,wang2023qhl} have been discussed in Section \ref{sec:Introduction} and they are all selfish routings on their own. One passive way to avoid selfish results is the \textit{alternative routing} \cite{abraham2013alternative,chondrogiannis2015alternative,kobitzsch2013alternative,li2021comparing,luxen2012candidate,lim2010efficient,chondrogiannis2018finding,hacker2021most,chondrogiannis2020finding,luo2022diversified,yu2020distributed} that provides several different routes for the same OD (Origin-Destination) with small overlapping while their lengths are still as short as possible. However, their choices (path candidates) are still limited without considering their actual traffic influence, and diversification has nothing to do with the traffic condition, so the effectiveness is limited. Another stream called \textit{continuous routing} \cite{malviya2011continuous,xu2012traffic,yang2014cands} aims to refine the routing results of the already on-road ones when traffic condition changes. It can improve the quality of the monitored routes, but still could harm traffic conditions especially when the monitored number increases due to its selfish nature. 

\textbf{2) Traffic Prediction}. 
Depending on the spatial granularity, the current traffic prediction can be categorized into i) \textit{OD}-based  \cite{li2018multi,yuan2020effective,wang2019simple} that estimates travel time of OD pairs without actual routes, ii) \textit{Path}-based \cite{zhang2018deeptravel,yang2013travel,hong2020heteta,hui2021trajnet} that estimates travel of the given routes, and iii) \textit{Road}-based \cite{wu2020connecting,zheng2020gman,pan2019urban,yu2018spatio,lv2018lc} that predict the travel time of each road in the network. They all depend on learning the spatiotemporal correlations \cite{wu2021autocts,cirstea2022towards,jin2023spatio,wu2023autocts+,zhang2023mlpst} from the historical data and require training process, so they can hardly react timely to ad-hoc changes of traffic distribution, accident and events, or topological changes. More importantly, our traffic optimization requires changing the routes, and it is essentially changing the historical patterns, which are the root of the current prediction paradigm. Finally, they can hardly scale to the large network.

\textbf{3) Traffic Simulation}. 
Traffic simulation tools like SUMO \cite{lopez2018microscopic, sumo2023official}, MATSim \cite{waraich2015performance, matsim2023official}, and SMARTS \cite{ramamohanarao2016smarts, xie2019generating, smarts2023official,ramamohanarao2017traffic} are widely used in the transportation research that mathematically models traffic by considering the traveling behaviors of vehicles like car-following, acceleration, line-changing, traffic lights, and turning. However, as accuracy is their priority and step-based simulation is inevitable, they are slow to run and can hardly support real-time optimization.

\textbf{4) Traffic Congestion Controlling}.
Most of these methods update edge weights from traffic influence but their usability is limited by predefined routes. i) \textit{Traffic Assignment} that assigns each trip into several alternative paths at each instant of time based on certain behavior rules (user-equilibrium) with game theories and operations research \cite{wardrop1952road, bell2002risk, friesz1993variational, gawron1998iterative, javani2019path,kachroo2013dynamic,szeto2012dynamic,perederieieva2015framework,kollias2021weighted,lim2012stochastic}, but they are very time-consuming and can only work on toy networks with a few nodes. ii) \textit{Penalty}-based method \cite{arp2020dynamic,chen2020pay} further discourages users from entering the congestion roads. iii) \textit{NextOpt} \cite{li2021towards} is the only work that integrates traffic influence into routing such that the weights are determined by the actual traffic flow, but it just optimizes the next road greedily so its effectiveness is limited.

\begin{figure*}[t!]
	\centering
	\includegraphics[width=6.8in]{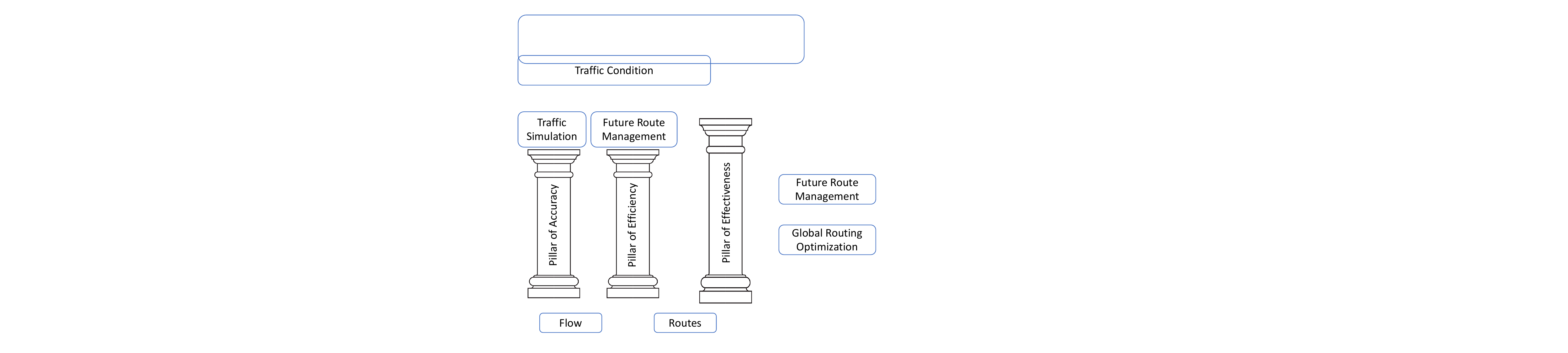}
	\caption{Next-Generation Routing System Framework}
	\label{fig:Framework}
\end{figure*}

\textbf{5) Trajectory Management}.
The routing result paths are a kind of trajectory, but they cannot be supported by the current trajectory databases \cite{zheng2015trajectory,zheng2011computing,wang2021survey,su2020survey} because their index structures are essentially static \cite{tao2001efficient,tao2001mv3r,tao2003tpr,ding2018ultraman,cudre2010trajstore,xie2017distributed,wang2015sharkdb} because the historical trajectories they managed seldom change. Although single-entry update is supported, they never consider one update's influence on the other data. Therefore, they cannot support the frequent dynamic route changes triggered by global optimization and cannot support efficient partial simulation.

\section{The New Routing Paradigm}
\label{sec:Paradigm}
In this section, we describe our new routing paradigm. Firstly, we emphasize that the root cause of traffic conditions is the actual traffic on a specific road rather than the historical record. Changing the actual route would change traffic conditions through the change of traffic flow. Therefore, managing traffic flow could generate more accurate traffic conditions, and changing traffic could provide opportunities towards global optimal. In addition, unlike other systems, the transportation system is relatively simple as the next state has fewer options (turning at intersections), and traffic rules and physical laws have fewer parameters. Therefore, we first define the basic concepts and then describe the framework.

\subsection{Preliminary}
\label{subsec:Paradigm_Preliminary}

\emph{\underline{Road Network}} is a graph $G=(V,E)$ with $V$ a set of road intersections and $E$ a set of road segments. 

\noindent\emph{\underline{Traffic Flow}} $f(u,v,t)$ is the number of vehicles on $(u,v) \in E$ at time $t$. When only a part of the total vehicles can be obtained, we regard the traffic flow as $f(u,v,t) = f_q(u,v,t) + f_o(u,v,t)$, where $f_q(u,v,t)$ the current system's, and $f_o(u,v,t)$ are the underlying ones. 

\noindent\emph{\underline {Route (Path)}} $r$ from $u$ to $v$ is represented as $r=\langle(v_0,t_0),\dots,$ $ (v_k,t_k)\rangle$, where $v_0 = u$, $v_k = v$, $(v_i, v_{i + 1})\in E$, and $t_i$ is the timestamp at $v_i$, $\forall 0 \leq i < k$. The travel time of $r$ departing at $t_0$ is defined as $T(r,t_0) = \sum_{i=0}^{k-1} c(v_i, v_{i+1},t_i)$, where $c(v_i, v_{i+1},t_i)$ indicates the travel time on $(v_i, v_{i+1}) \in E$ at time stamp $t_i$.

\noindent\emph{\underline {Query}} $q(u,v,t)$ asks for a path $r$ from $u$ to $v$ departing at time $t$. The query results are used to represent the traveling traffic and opportunities for optimization.

\noindent\emph{\underline{Latency Function / Model}}. For each edge $(u,v)\in E$, the travel time $c(u,v,t)$ varies dynamically based on the traffic flow $f(u,v,t)$ and other static road features such as road type, number of lanes, speed limit, traffic light, turnings, weather, and etc. 

\noindent\emph{\underline{Traffic-Aware Road Network}} is a road network whose edge weights are determined by the latency function.  

\textbf{Next Generation Routing System Requirement:} Given a traffic-aware road network and a continuous set of queries $Q=\{q_i\}$ during a time period, the routing system should optimize the routes of $Q$ such that $\sum_{q_i\in Q}T(r_i,t_i)$ is as small as possible.

\subsection{Next-Gen Routing System Framework}
\label{subsec:Paradigm_Framework}
In this part, we present our next-generation routing system with the illustration from Figure \ref{fig:Framework}. Firstly, we view all the travels as routing queries and their routes are planned with the traffic conditions and used for navigation. At this stage, it is the same as the existing routing paradigm, and its selfishness would cause congestion. After that, routes are converted into traffic flows and fed into traffic simulation to generate traffic conditions, and the finished ones are archived into trajectory databases. Until now, it is the same as the existing traffic simulation system, and its inefficiency is unacceptable for real-time scenarios. As a result, the routing system still has to use the predicted ones rather than the simulated ones, no matter how accurate they are.

What the next-generation routing system different from the existing ones are 1) its iterative nature caused by traffic conditions and routing results intertwined with each other like chicken and eggs, and 2) the replacement of the three components with three new ``pillars'' with extra requirements. In the following, we describe the routing system from the perspective of procedures such that the pillars' roles are identified within the iterations, as illustrated in Figure \ref{fig:Framework}-(b). 

In the first round, the queries are fed into the \textit{routing optimization} to generate the first batch of routes. This part could utilize the existing pathfinding solutions or distribute the results passively based on the road features. 
Then the routes are regarded as traffic flow for traffic simulation. As the first-round routing has no traffic in the network, we could view it as a traditional \textit{simulation} task. But its efficiency has to improve to support later iterations, so it has to change from the accurate time-step-based \textit{microscopic} simulation to the approximate road-based \textit{macroscopic} simulation. At the same time, these simulation results are fed into the \textit{route management system} to support future simulation. When the simulation finishes, the traffic conditions caused by this batch of queries are obtained, and we can either plan the routes for the next batch of queries or re-plan some of the current ones to improve the traffic condition. It should be noted that as long as the vehicles are still traveling, their future routes can still be optimized continuously. 

In the following rounds, the \textit{global routing optimization} plan the routes according to the latest traffic conditions, the newly generated routes are inserted into the \textit{future route management} such that \textit{traffic simulation} can be updated incrementally based on the previous simulation results but not from ground-up. In addition, for optimizing the existing routes, the old re-routed ones could be deleted from the \textit{future route management}, and the newly updated routes are inserted for simulation. In this way, the influence of the new queries and the adjusted ones generate new traffic conditions in real-time for the next rounds on and on.

Within this framework, the \textit{traffic simulation} is to guarantee the effectiveness of the simulation, the \textit{future route management} is to support efficient simulation, and \textit{global routing optimization} determines the actual routes based on the simulation results for effective traffic condition optimization. In the following three sections, we will elaborate on the expectations, preliminary design, and experimental results of these three pillars.
 
\subsection{Data Dilemma: Ground Truth \textit{vs} Simulation}
\label{subsec:Paradigm_Data}
One big obstacle to global routing research (and all the related research of traffic forecasting, trajectory management, spatiotemporal analysis, \etc), is the limited access and amount to the real-life ground-truth data, which often causes the questioning on the practicability of the research output. In this section, we identify that the ``practicability'' is actually another layer of the problem and explain why our research should not be limited by it. 

Firstly, the traffic has to travel according to the physical laws and traffic rules. The detailed behavior and accuracy are irrelevant to the downstream research, and they are the topics of the transportation area (upper of Figure \ref{fig:Framework}-(c)). In fact, the traffic simulation tools require tuning the parameters to reflect different situations in real life. In this way, when the parameters change, it can simulate a new situation. In other words, it has the capability to generate all kinds of traffic situations even the ones that have not happened before.

Secondly, the downstream research should have the capability to handle all situations but not be limited to the existing real-life data. For instance, in our problem, the future traffic patterns might be caused by a new scenario that has never happened in real life. Then it would be impossible to compare with the ground truth. Moreover, the simulation system could evaluate any outputs. Therefore, from our perspective, we can regard the existing simulation system as ground truth, and our latter system's accuracy is evaluated based on it (lower part of Figure \ref{fig:Framework}-(c)). What is worse, forcing the use of real-life data, which is hard to collect and publish due to privacy issues, could harm future research. This is because most of them (like PeMS datasets \cite{PEMS2023official}) are rather small with hundreds of roads and belong to the same type like highways, which limits their effectiveness only to these datasets and hardly applied to other larger, more complicated, and unseen situations. Although synthetic data could be added to simulate new scenarios, their validity is highly questionable as they are not generated from proper simulation.

Therefore, we believe that only by utilizing the simulation tools can the generality, scalability, and optimization research proceed, so our research on the next-generation routing system is built and evaluated upon traffic simulation data instead of real-life data. Besides, providing a high-quality benchmark simulation dataset that covers different scenarios and supports different analysis tasks would also benefit the whole research community. As the data are purely simulated, it would be free from privacy concerns and free for everybody to use and compare.

\section{The Pillar of Accuracy: Traffic Condition Simulation}
\label{sec:Simulation}
traffIn this section, we present the first component, the traffic condition simulation part, which guarantees the accuracy of the whole system. Without it, the optimization result would become untrustworthy, so we call it the \textit{Pillar of Accuracy}. The main idea is to replace the time-step-based microscopic simulation which considers each vehicle's detailed movement, with road-based macroscopic simulation which simplifies the vehicles into aggregated traffic flow, to improve the efficiency. The main issue is how to retain the simulation quality. In the following, we discuss the mapping from traffic flow to traffic condition from the road level to the network level.

\subsection{From Traffic Flow to Travel Time}
\label{subsec:Simulation_Flow}
Given a traffic flow and a road with its features, the unit task is how to generate the corresponding travel time. The first solution could be utilizing the \textit{Bureau of Public Roads (BPR)} \cite{lim2012stochastic, manual1964urban} that is widely used in transportation engineering:
\begin{equation}
\label{eq:BPR}
    c(u,v,t) = c_{min}(u,v) \times (1+\sigma \times {(\frac{f(u,v,t)}{\varphi})}^{\beta}) 
\end{equation}
$c_{min}(u,v) = \frac{road \ length}{limited \ speed}$ is the free-flow ($f(u,v,t)=0$) or minimum travel time on the road $(u,v)$, $\varphi$ is the threshold road capacity and $(\frac{f(u,v,t)}{\varphi})$ is the percentage of road occupation, the power $\beta$ varying from 2 to 4 is the penalty when the flow surpasses the threshold, and $\sigma$ is the weight of the penalty. The settings of these parameters differ from road to road and from time to time. The current solution is setting them from expert experience, which can hardly react to any dynamic changes. Therefore, given enough simulated data of each road in different scenarios, we can train a model to map the current road features to these parameters. Another choice is discarding the \textit{BPR} function and its parameters and training a model directly from features to travel time. 

%
%

\subsection{Macroscopic Traffic Simulation}
\label{subsec:Simulation_Simulation}
In this section, we describe how to use the traffic model for macroscopic simulation. Firstly, we assume that when a vehicle enters a road $(u,v)$ at time $t$, it is the traffic flow $f(u,v,t)$ that determines the actual time to go through $(u,v)$. In addition, as the vehicles entering after $t$ are behind the current vehicle, they would not influence its travel time. Therefore, we only need to visit the vertices along all the paths in a time-increasing order to determine each path's next road travel time. Specifically, we use a label $(r,i,t)$ to record the route $r$ has arrived at its $i^{th}$ vertex at time $t$. Then we use a heap $L$ to organize all routes' labels in their departure time-increasing order. Each time we retrieve the earliest label $l$ from $L$, remove it from the previous road and decrease its flow by 1, and move it forward to the next road and increase its flow by 1. If $l$ has reached its destination, then accumulate its travel time to $T(r)$. Otherwise, $l$ is pushed back to $L$. When all the paths have finished their traversal, we have obtained the total travel time and their detailed schedules. This procedure has a complexity of  $O(|V||P|\log|P|)$. Its correctness and detailed procedures can be found in \cite{xu2023global}.

\begin{figure*}[t!]
	\centering
	\includegraphics[width=6.8in]{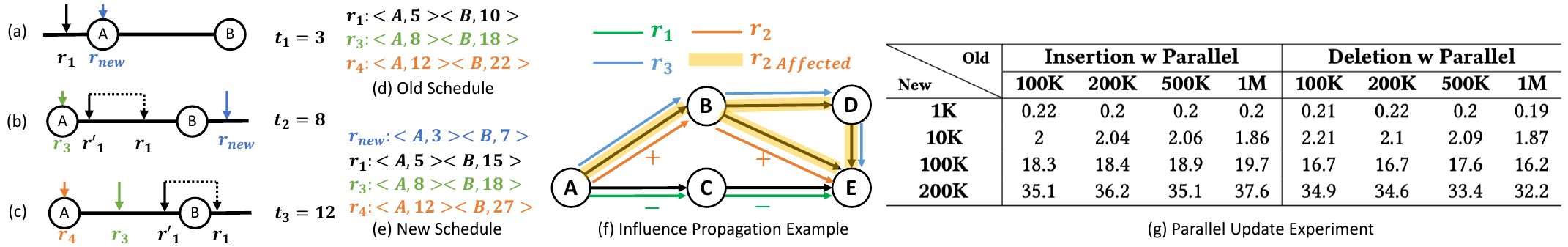}
	\caption{Future Route Management Examples and Preliminary Experiment Results}
	\label{fig:Simulation}
\end{figure*}

\subsection{From One Large Model to Managing Several Small Models}
\label{subsec:Simulation_DB4Simulation}
In either of the two ways described in Section \ref{subsec:Simulation_Flow}, we face a smaller task of one single road than the ordinary traffic predictions over the entire network, so their accuracy and efficiency are expected to be better on the road level. However, such a strategy still faces the following issues: 1) The actual routes are made up of multiple consecutive roads, and simply adding the travel time generated from each road together cannot guarantee a correct travel time. Besides, for the routes traveling on the same road, they came from different previous roads and would go to different next roads, so their travel time might be different due to the traffic light and the waiting queue; 2) As each road corresponds to a model, there would $|E|$ models to manage, which is not practical in real-life; 3) As a unit task that would be called $|r_i|$ times for each route and as large as $\sum_{i\in[0,|Q|)}|r_i|$ times for one query batch, its efficiency has to be as fast as possible. Therefore, we need the following three tasks to solve these issues.

\textbf{1) Route-Incorporated Model:} Essentially, our traffic model should be able to consider the influence of traffic lights on the different routes. To this end, the travel time to different neighboring roads should be different, so the next direction of a vehicle should also be an input feature when computing its travel time. Meanwhile, as different directions could be waiting in different lanes and have different lengths of passing time, the traffic flow itself should also be categorized based on directions. Moreover, when training the models, they should not be treated as standalone ones but should be linked together to form small regional networks such that the continuous influence of the simulated route data could be captured. The whole training process could follow the structure described in Section \ref{subsec:Simulation_Simulation} in a DB4AI fashion with the microscopic results as ground truth. In this way, it could be expected the macroscopic simulation result would be close to the microscopic.

\textbf{2) Model Management:} On one hand, one specific model for a road can be lightweight and fast to use, but there is no need to train so many of them. On the other hand, a larger model has better generalizability and is easy to manage, but it is harder to train, slower to run, and less accurate for specific tasks than the small models. Therefore, we propose to use several smaller models for different types of roads. This task would involve traffic-based road clustering and model selection. 

\textbf{3) Materialization:} To further improve this basic unit's efficiency, we could further materialize the model down to flow-time mapping when the result has little variation. This part requires a detailed analysis and classification of each road's traffic model and identify the opportunity to materialize as much as possible.

\section{The Pillar of Efficiency: Future Route Management}
\label{sec:Management}
In this section, we present the second component, the future route management part, which guarantees the efficiency of the whole system. Without it, the simulation and optimization results would be easily outdated and cannot apply to real life, so we call it the \textit{Pillar of Efficiency}. The main idea is to enable a new ``traffic-aware update'' operation to the trajectory database, and the issue is how to retain high efficiency to support frequent updates as the optimization process needs its result for evaluation. Another difference is the scope: because only the near future changes frequently while the historical ones are static, the current should be monitored and a future window should be simulated. Therefore, this system is also a backbone infrastructure for the urban city digital twin.

\subsection{Future Route Index and Management}
\label{subsec:Management_Index}
Essentially, we need to manage the simulation results generated by Section \ref{subsec:Simulation_Simulation} such that the updated locations and influenced routes can be identified efficiently. As routes are all network-constrained, we can use a graph to organize them. Apart from the original routes information, because the flows are from each road's perspective, we re-organize the routes into different roads they traversed (like an inverted list), and the in-/out- flow changes caused by them are indexed with a B-Tree in the time-increasing order within each road. In other words, we are managing the flow changes. 

\subsection{Influence Identification and Propagation}
\label{subsec:Management_Influence}
In the previous flow-traffic mapping assumption, when a route is inserted or deleted from a road, it changes the flow, which further influences the future routes recursively on the current road. What's worse, the influenced routes would further influence its future roads, so the influence would also propagate spatially. Therefore, this component requires the following parts:

\textbf{1) Single-Road Influence Termination:}  For example in Figure \ref{fig:Simulation}-(a), a new route $r_{new}$ enters road $(A,B)$ before $r_1$. In (d) and (e), the old schedule is affected, and $r_1$ becomes the first affected route that travels slower and leaves $(A,B)$ later from 10 to 15. Then, in (b), when $r_3$ arrives at $A$, $r_1$ is still on $(A,B)$ (but delayed to $r'_1$), and its travel time remains the same as in the old schedule because the traffic flow is still 1. Therefore, $r_3$ becomes the first unaffected route. However, this is not the end of the influence scope, as $r_4$ is affected by this change as well. As (c) shows, in the old schedule, the traffic flow was 1 when $r_4$ arrived at $A$, but now with $r_1$ traveling slower (delayed to $r'_1$), the traffic flow becomes 2, and $r_4$ travels slower as a result. So this task needs to determine the first unaffected route. 

\textbf{2) Multi-Road Propagation:} In Figure \ref{fig:Simulation}-(f), $r_1=\langle A,C,E\rangle$. However, if it reroutes to $r_2=\langle A,B,E\rangle$, then the previous simulation becomes incorrect. Consequently, route $r_1$ is deleted, and a new route $r_2$ is added to re-simulate. After that, $r_3=\langle A,B,D,E\rangle$'s travel time may also increase due to the change of $(A,B)$, which could further propagate to roads $(B,D)$ and $(D,E)$ (roads highlighted in yellow are affected by this re-route). In summary, we need to identify the affected roads and routes, and then organize them in some order to avoid repeated maintenance. 

\begin{figure*}[t!]
	\centering
	\includegraphics[width=6.8in]{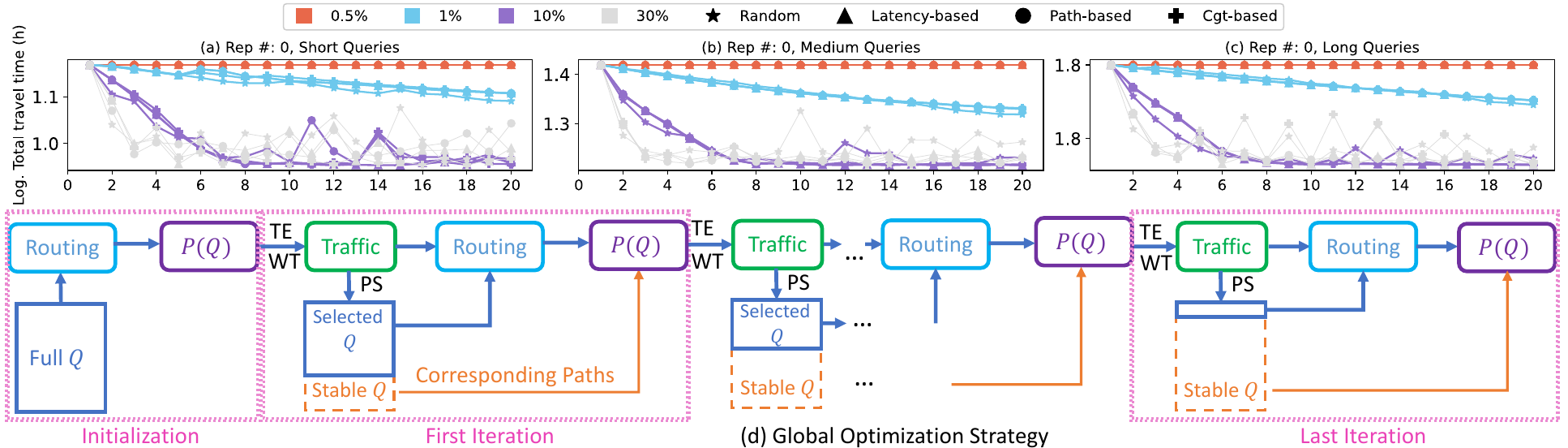}
	\caption{Global Optimization Strategy and Preliminary Experiment Results}
	\label{fig:Optimization}
\end{figure*}

\textbf{3) Multi-Route Propagation:} When we optimize the routes, it is common to have several deletions and insertions occur at the same time. Therefore, we first prove the simulation update operations could be conducted in parallel, and then test it on the Beijing network with 96,710 vertices and 774,660 edges. The results are shown in Figure \ref{fig:Simulation}-(g), with columns are the stored routes and rows are the updated ones. In the largest case, our system can re-simulate 200k new routes with 1 million existing routes around 30s and 1k new routes within 1s, while the total re-simulation takes several minutes regardless of the update number.



\section{The Pillar of Effectiveness: Global Routing Optimization}
\label{sec:Optimization}
In this section, we present the third component, the global routing optimization part, which guarantees the effectiveness of congestion easement. Without it, the routing results are still selfish, so we call it the \textit{Pillar of Effectiveness}. The main idea is to distribute the routes in the initial routing phase and then re-route the congested ones.

\subsection{Diversified Routing}
\label{subsec:Optimization_Diversified}
The existing methods only take the length of overlapped routes as the similarity function, so the results are not optimized for traffic conditions and can hardly reduce congestion. Therefore, a new similarity evaluation among paths that considers traffic conditions is required, which further needs to incorporate time dependency, road features, and traffic flow. Meanwhile, computational efficiency is still a bottleneck especially when routes are long, network is dense, result number is large, and similarity threshold is low. 

\subsection{Optimization Strategies}
\label{subsec:Optimization_Strategies}
Firstly, the optimization works in an iterative way as shown in Figure \ref{fig:Optimization}-(d), where we select a sub-set of congested queries to re-route based on the previous simulated traffic conditions. Such a procedure is analogous to the \textit{gradient descent} optimization method, where the traffic condition is the gradient (lead the re-routing towards the faster / uncongested roads), the number of re-routing queries is the step size, and the simulation result is the error. The optimization goal is the total simulated travel time of all the routes. As described in \cite{xu2023global}, there are several strategies (\textit{random, latency, path, congestion}) to selecting queries to re-route, and the percentage of selection reflects the re-route number. Figure \ref{fig:Optimization}-(a) to (c) shows the optimization results on Beijing road networks of the short, medium, and long queries with the combination of the above parameters. Intuitively, iteration 0 is the selfish routing, which has the highest total travel time. As the iteration number increases, the global travel time decreases as expected. Among them, 0.5\% and 1\% decrease slowly while 30\% decreases the fastest but is very unstable. 10\% achieves the balance between stability and efficiency and can reach optimal after 8 to 10 rounds. As for the selection criteria, random is the most unstable while the other three have similar performance. More experimental results, strategies, and explanations can be found in \cite{xu2023global}.

The above re-routing still assumes the departure time is the same as the original one, but its effectiveness would drop when there are so many queries such that congestion is inevitable no matter how to schedule them. To this end, distributing the re-route queries' departure time with the \textit{interval-departure time fastest path algorithm} \cite{kanoulas2006finding,ding2008finding,li2017minimal} could further reduce the congestion. Initial results show that the fixed departure setting is effective when we repeat the query number under 4 times, while the flexible departure version could tolerate repeat number of 50 times. 

Re-routing efficiency is another important aspect as it requires fast routing on large dynamic and time-dependent networks. Currently, the re-routing is implemented through \textit{Dijkstra}'s search with multi-threads, while none of the existing path indexes could contribute to this problem because of the high update frequency and volume triggered by the simulation and high complexity of the time-dependency.

\section{Conclusion}
\label{sec:Conclusion}
In this vision paper, we describe the next-generation routing system that combines the routing and evaluation together for real-time transportation optimization, and then introduce the three key components to support it: \textit{traffic simulation} for accuracy, \textit{future route management} for efficiency, and \textit{global optimization} for effectiveness. Each sub-system introduces a set of new problems that require lots of effort to solve. Since their performance is intertwined with each other, we will fulfill the system's overall optimization promise when all three pillars are ready.
Therefore, we appeal to our research community to focus more on each pillar's own performance at this early stage and postpone the practical validation to later on.




\balance
\bibliographystyle{unsrt}
  \bibliography{References}

\end{document}